\documentclass[twocolumn,paper]{jpsj2}
\usepackage{color}

\def\runauthor{Akira \textsc{Oosawa} {\it et al.}}

\title{Polarized Neutron Inelastic Scattering Study of the Anisotropic Magnetic Fluctuations in the Quasi-1D Ising-like Antiferromagnet TlCoCl$_3$}

\author{Akira \textsc{Oosawa}\thanks{E-mail address: osawa-a@sophia.ac.jp}, Kazuhisa \textsc{Kakurai}$^{1}$, Yoichi \textsc{Nishiwaki}$^{2}$ and Tetsuya \textsc{Kato}$^{3}$}

\inst{Department of Physics, Sophia University, 7-1 Kioi-cho, Chiyoda-ku, Tokyo 102-8554, Japan\\
$^1$Quantum Beam Science Directorate, Japan Atomic Energy Agency, Tokai, Ibaraki 319-1195, Japan\\
$^2$Department of Physics, Tokyo Institute of Technology, Oh-okayama, Meguro-ku, Tokyo 152-8551, Japan\\
$^3$Faculty of Education, Chiba University, 1-33 Yayoi-cho, Inage-ku, Chiba 273-8522, Japan}

\recdate{\today}

\abst{Polarized neutron inelastic scattering experiments have been carried out in the quasi-1D Ising-like antiferromagnet TlCoCl$_3$. We observed the longitudinal magnetic fluctuation $S_{zz} ({\pmb Q}, \omega)$ for the spin-wave excitation continuum, which has not been observed in the unpolarized neutron inelastic scattering experiments of the quasi-1D Ising-like antiferromagnets CsCoCl$_3$ and TlCoCl$_3$ so far, together with the transverse magnetic fluctuation $S_{xx} ({\pmb Q}, \omega)$. We compared both obtained intensities of $S_{xx} ({\pmb Q}, \omega)$ and $S_{zz} ({\pmb Q}, \omega)$ with the perturbation theory from the pure Ising limit by Ishimura and Shiba (IS), and a semi-quantitative agreement was found.}  

\kword{TlCoCl$_3$, quasi-1D Ising-like antiferromagnet, domain wall, excitation continuum, anisotropic magnetic fluctuations, polarized neutron inelastic scattering}

\begin{document}
\sloppy
\maketitle

\section{Introduction}

The magnetic excitations of the 1D Ising-like antiferromagnet with the following Hamiltonian,
\begin{equation}
{\cal H} = 2J \sum_j \left[ S_j^z S_{j+1}^z + \epsilon \left( S_j^x S_{j+1}^x + S_j^y S_{j+1}^y\right) \right]
\end{equation}
have been attracting much attention because the excitations are characterized as the domain-wall excitations and the behavior of the excitations is much different from that of the conventional spin-wave excitations of the Heisenberg antiferromagnet. \par
Previously, the unpolarized neutron inelastic scattering experiments have been performed on the quasi-1D Ising-like antiferromagnet CsCoCl$_3$ \cite{Yoshizawa,Nagler}. In the experiments, two magnetic excitations, which are the central mode corresponding to the motion of thermally activated domain walls firstly predicted by Villain \cite{Villain} and the spin-wave excitation continuum corresoponding to the domain-wall pair excitation, were observed. For the central mode, both the transverse magnetic fluctuation $S_{xx} ({\pmb Q}, \omega)$ and the longitudinal magnetic fluctuation $S_{zz} ({\pmb Q}, \omega)$ were observed and the separation of the two fluctuations has been successfully performed using the property of the neutron magnetic inelastic scattering cross section, namely the magnetic fluctuations perpendicular to the scattering vector ${\pmb Q}$ is only contributing to the cross section. While, for the spin-wave excitation continuum, the separation of both fluctuations failed because the intensity of $S_{zz} ({\pmb Q}, \omega)$ is much smaller than that of $S_{xx} ({\pmb Q}, \omega)$ so that $S_{zz} ({\pmb Q}, \omega)$ is masked by $S_{xx} ({\pmb Q}, \omega)$ out of the scattering plane which is always observed in the unpolarized experiments. \par
In this paper, we demonstrate the advantages of the polarized neutron inelastic scattering experiments for the investigation of the anisotropic magnetic fluctuations in the 1D Ising-like antiferromagnet, which enables the separation of $S_{xx} ({\pmb Q}, \omega)$ and $S_{zz} ({\pmb Q}, \omega)$, and report the polarized neutron inelastic scattering investigation on the quasi-1D Ising-like antiferromagnet TlCoCl$_3$. \par
TlCoCl$_3$ has the CsNiCl$_3$-type hexagonal crystal structure (space group symmetry $P6_3/mmc$) at room temperature, in which the magnetic Co ions form the linear chain along the $c$-axis and these chains make triangular lattice in the $c$-plane. Because Co ions have a large Ising anisotropy along the $c$-axis due to the Kramers doublet, this system can be expressed as the quasi-1D Ising-like antiferromagnet. From the dielectric constant measurements, it was found that this system undergoes the successive structural phase transitions with ferroelectricity at $T_{\rm st2}=165$~K, $T_{\rm st3}=75$~K and $T_{\rm st4}=68$~K \cite{Nishiwaki1}. Besides, it was found from the magnetization measurements that the magnetic phase transition occurs at $T_{\rm N}$=29.5~K \cite{Nishiwaki1}. The neutron diffraction measurements have been carried out in this system in order to investigate the crystal and magnetic structures. It was found that the crystal structure varies from the hexagonal structure ($P6_3/mmc$) at room tempearture to the orthorombic one ($Pbca$) below $T_{\rm st4}$ and the magnetic Bragg reflections indicative of the {\it up-up-down-down}-type magnetic structure have been observed below $T_{\rm N}$ at ${\pmb Q}=(\frac{h}{8}, \frac{h}{8}, l)$ with odd $h$ and odd $l$, which are assigned by the indices of the room-temperature hexagonal lattice \cite{Nishiwaki2}. The unpolarized neutron inelastic scattering measurements have been performed in order to investigate the magnetic excitations in TlCoCl$_3$ \cite{TlCoCl3inela}. The spin-wave excitation continuum was observed. According to the IS theory \cite{Ishimura}, we analyzed the observed spin-wave excitation, and the exchange constant $2J$ and the anistropy $\epsilon$ were estimated as 14.7~meV and 0.14 in TlCoCl$_3$, respectively.
\begin{figure}[t]
\begin{center}
\includegraphics[width=85mm]{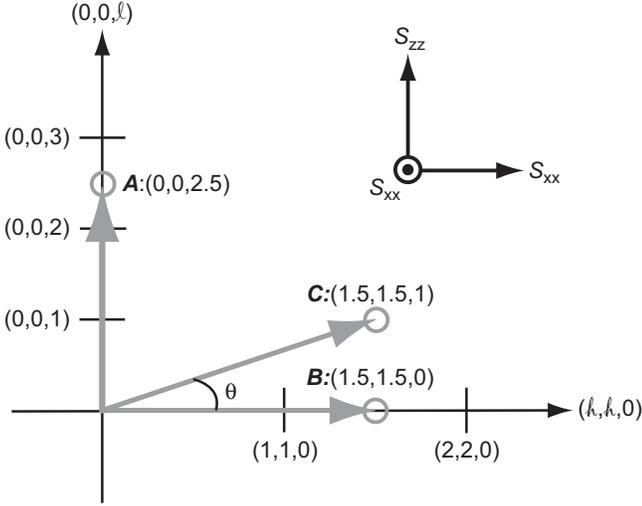}%
\end{center}
\caption{Scattering plane of the present experiment. The gray arrows denote the scattering vectors ${\pmb Q}$ in the present measurements. $\theta$ is the angle between the scattering vector ${\pmb Q}$ and $(h, h, 0)$-axis. The magnetic fluctuations along each directions are also shown. \label{Fig1}}
\end{figure}

\section{Experimental Details}

Polarized neutron inelastic scattering experiments using the uniaxial polarized neutron scattering technique \cite{Moon} were performed using JAEA-TAS1 spectrometer installed at JRR-3M in Tokai. The constant-$k_{\rm f}$ mode was taken with a fixed final neutron energy $E_{\rm f}$ of 14.7~meV. Heusler(111)-Heusler(111) monochromator-analyser configuration was used in the present experiments. The polarization was measured on the nuclear Bragg reflection and was determined to be around 96~\%. A pyrolitic graphite-filter was used to suppress the higher order contaminations. The collimations were set as open-80'-open-open in order to gain intensity. We used a single crystal of TlCoCl$_3$ with a volume of approximately 0.5~cm$^3$. The present measurements were performed at $T=34$~K, which is just above $T_{\rm N}$. The crystallographic parameters were determined as $a = 6.84$~${\rm \AA}$ and $c = 5.93$~${\rm \AA}$ at $T=34$~K, which are defined in the room-temperature hexagonal lattice. The sample was mounted in the cryostat with its $(h, h, 0)$- and $(0, 0, l)$-axes in the scattering plane, as shown in Fig. \ref{Fig1}. The indices of the room-temperature hexagonal lattice are used in the present experiments. The relation between the indices of the room-temperature hexagonal lattice and those of the low-temperature orthogonal lattice is shown in ref. \citen{Nishiwaki2}. In the present experiments, the guide field was applied perpendicular to the scattering $(h, h, l)$ plane, {\it i.e. } vertical field configuration. In this configuration, the inelastic magnetic scattering intensity in the spin-flip (SF) channel is proportional to the intensity of the magnetic fluctuations perpendicular to the vertical guide field, {\it i.e.} fluctuations in the scattering plane, while that of the non-spin-flip (NSF) channel is proportional to the intensity of the magnetic fluctuations parallel to the guide field, namely fluctuations out of the scattering plane as follows:
\begin{eqnarray}
\label{A}
I_{\rm NSF} &=& C \cdot S_{xx} ({\pmb Q}, \omega) \\
\label{AA}
I_{\rm SF} &=& C \left( \sin^2 \theta \cdot S_{xx} ({\pmb Q}, \omega) + \cos^2 \theta \cdot S_{zz} ({\pmb Q}, \omega) \right) \nonumber \\
\end{eqnarray}
where $C$ and $\theta$ are the scale factor and the angle between the scattering vector ${\pmb Q}$ and $(h, h, 0)$-axis, respectively. Hence the longitudinal fluctuation $S_{zz} ({\pmb Q}, \omega)$ can be separated from the transverse fluctuation $S_{xx} ({\pmb Q}, \omega)$ by measurements at a single $({\pmb Q}, \omega)$ point.
\begin{figure}[t]
\begin{center}
\includegraphics[width=85mm]{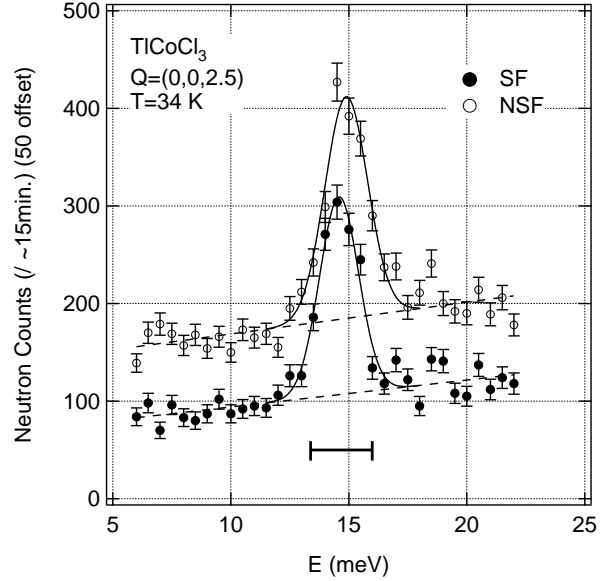}%
\end{center}
\caption{Profiles of the constant-${\pmb Q}$ energy scans in TlCoCl$_3$ for NSF and SF scatterings at ${\pmb Q}$ = (0, 0, 2.5). The solid and dashed lines denote the fits by a Gaussian function and the energy dependent backgrounds, respectively. The horizontal bar indicates the calculated resolution width at $E=14.7$ meV. \label{Fig2}}
\end{figure}

\section{Results and Discussion}

First, we measured the NSF and SF scattering intensities for ${\pmb Q}=(0, 0, 2.5)$ (Point-A in Fig. \ref{Fig1} ). Because $\theta$ is equal to 90$^{\circ}$ at ${\pmb Q}=(0, 0, 2.5)$, we can expect that the same intensities, {\it i.e.} $C \cdot S_{xx} ({\pmb Q}, \omega)$, are observed for both NSF and SF scatterings. Figure \ref{Fig2} shows the profiles of the constant-${\pmb Q}$ energy scans in TlCoCl$_3$ for NSF and SF scatterings at ${\pmb Q}$ = (0, 0, 2.5). As shown in Fig. \ref{Fig2}, the excitations with resolution limited and same intensities are clearly seen for both scatterings. From the IS theory, it was found that $S_{xx} ({\pmb Q}, \omega)$ in the 1D Ising-like antiferromagnet can be expressed as the following expression for the peak position
\begin{equation}
\label{B} 
\omega_q = 2J (1 - 8 \epsilon^2 \cos^2 \pi l)
\end{equation}
and the boundaries of the continuum extending from $\omega_q^-$ to $\omega_q^+$, given by
\begin{equation}
\label{C}
\omega_q^{\pm} = 2J (1 \pm 2 \epsilon \cos \pi l).
\end{equation}
Hence, at ${\pmb Q}$ = (0, 0, 2.5), the continuum should be closed so that the magnetic excitation with resolution limited should be observed at $\omega=2J$. Because the exchange constant $2J$ has been already estimated as 14.7 meV in this system \cite{TlCoCl3inela}, the observed excitations at ${\pmb Q}=(0, 0, 2.5)$ are consistent with the IS theory. Note that this system undergoes the successive structural phase transitions from the hexagonal crystal structure to orthorombic one \cite{Nishiwaki1, Nishiwaki2}. However, it was found that structural phase transitions scarcely affect $S_{xx} ({\pmb Q}, \omega)$ because the magnetic excitations with same intensities were observed for both scatterings. The sloping background in Fig. \ref{Fig2} can be ascribed to the transfer energy dependent counting time due to the chosen constant-$k_{\rm f}$ mode, where the incident neutron energy is varied while the outgoing neutron energy is fixed at $E_{\rm f}=14.7$ meV. Since we measure normalized on the incident beam monitor counts, the counting time increases as the transfer energy is increased and concomitantly the instrumental background increases. These ${\pmb Q}$-independent sloping backgrounds for both channels will be fixed in all the following profiles to compare the observed intensities with the calculation. \par
\begin{figure}[t]
\begin{center}
\includegraphics[width=85mm]{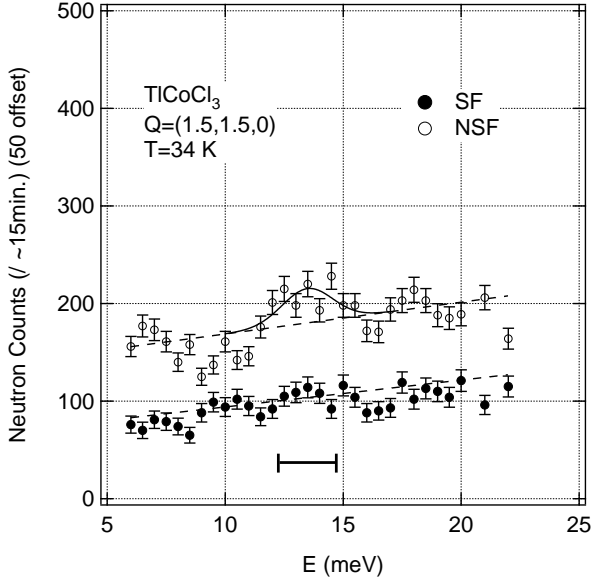}%
\end{center}
\caption{Profiles of the constant-${\pmb Q}$ energy scans in TlCoCl$_3$ for NSF and SF scatterings at ${\pmb Q}$ = (1.5, 1.5, 0). The solid and dashed lines denote the guide for the eyes and the energy dependent backgrounds estimated from the scan profiles at ${\pmb Q}$ = (0, 0, 2.5), as shown in Fig. \ref{Fig2}, respectively. The horizontal bar indicates the calculated resolution width at $E=13.5$ meV. \label{Fig3}}
\end{figure}
Next, we measured the NSF and SF scattering intensities for ${\pmb Q}=(1.5, 1.5, 0)$ (Point-B in Fig. \ref{Fig1} ). Because $\theta$ is equal to 0$^{\circ}$ at ${\pmb Q}=(1.5, 1.5, 0)$, we can expect from eqs. (\ref{A}) and (\ref{AA}) that only $S_{xx} ({\pmb Q}, \omega)$ and $S_{zz} ({\pmb Q}, \omega)$ are observed for NSF and SF scatterings, respectively. Figure \ref{Fig3} shows the profiles of the constant-${\pmb Q}$ energy scans in TlCoCl$_3$ for NSF and SF scatterings at ${\pmb Q}$ = (1.5, 1.5, 0). As shown in Fig. \ref{Fig3}, no excitation can be seen for the SF scattering while a little peak structure is seen for the NSF scattering. In the IS theory, the intensity of $S_{zz} ({\pmb Q}, \omega)$ can be expressed as
\begin{equation}
\label{D}
S_{zz} ({\pmb Q}, \omega) = \epsilon^2 \sin^2 \frac{\pi l}{2} \times \frac{1}{\pi |V|} \sqrt{1 - \left( \frac{\omega - 2J}{2 |V|} \right)^2}
\end{equation}
where
\begin{equation}
V = \epsilon J (1 + e^{-2 \pi i l} )
\end{equation}
while that of $S_{xx} ({\pmb Q}, \omega)$ can be expressed as 
\begin{equation}
\label{E}
S_{xx} ({\pmb Q}, \omega) \simeq \left\{ \begin{array}{l}
                                        \frac{1}{8 \pi |V|^2} \sqrt{4 |V|^2 - \left( \omega - 2J \right)^2} \left[ 1 - 2 \epsilon \cos \pi l \right. \\ 
                                        \hspace{0.5cm}  \left. - \frac{\omega - 2J}{J} \right] \hspace{0.5cm} \mbox{for} \hspace{0.5cm} |\omega - 2J| < 2|V| \\
                                        0 \hspace{0.5cm} \mbox{otherwise.}
                                        \end{array} \right.                                      
\end{equation}
As shown in eqs. (\ref{D}) and (\ref{E}), $S_{zz} ({\pmb Q}, \omega)$ should become 0 while $S_{xx} ({\pmb Q}, \omega)$ is finite when $|\omega - 2J| < 2|V| = 4 \epsilon J$ at ${\pmb Q}$ = (1.5, 1.5, 0). This prediction is consistent with the experimental results, as shown in Fig. \ref{Fig3}. \par
\begin{figure}[t]
\begin{center}
\includegraphics[width=85mm]{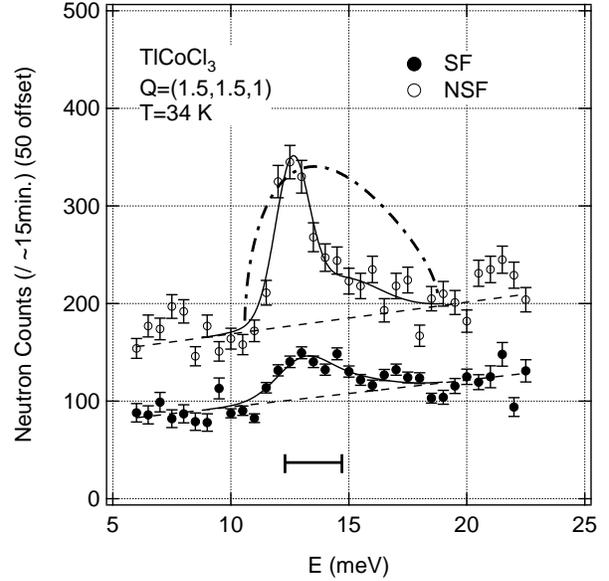}%
\end{center}
\caption{Profiles of the constant-${\pmb Q}$ energy scans in TlCoCl$_3$ for NSF and SF scatterings at ${\pmb Q}$ = (1.5, 1.5, 1). The solid and dashed lines denote the guides for the eyes and the energy dependent backgrounds estimated from the scan profiles at ${\pmb Q}$ = (0, 0, 2.5), as shown in Fig. \ref{Fig2}, respectively. The dash-dotted line denotes the NSF scattering intensity calculated from the $S_{xx} ({\pmb Q}, \omega)$ spectrum of the IS theory (eq. (\ref{F}) ) with the appropriate parameters given in the text. The horizontal bar indicates the calculated resolution width at $E=13.5$ meV. \label{Fig4}}
\end{figure}
In order to detect $S_{zz} ({\pmb Q}, \omega)$, we measured the NSF and SF scattering intensities for ${\pmb Q}=(1.5, 1.5, 1)$ (Point-C in Fig. \ref{Fig1} ). Because $\theta$ is estimated as 21$^{\circ}$ at ${\pmb Q}=(1.5, 1.5, 1)$, we can expect from eqs. (\ref{A}) and (\ref{AA}) that the intensities corresponding to $S_{xx} ({\pmb Q}, \omega)$ and $0.13 S_{xx} ({\pmb Q}, \omega) + 0.87 S_{zz} ({\pmb Q}, \omega)$ are observed for NSF and SF scatterings, respectively. Figure \ref{Fig4} shows the profiles of the constant-${\pmb Q}$ energy scans in TlCoCl$_3$ for NSF and SF scatterings at ${\pmb Q}$ = (1.5, 1.5, 1). In the NSF scattering, we observed the magnetic excitation with asymmetric spectral shape which is the same as that observed in the unpolarized experiments of CsCoCl$_3$ \cite{Yoshizawa,Nagler} and TlCoCl$_3$ \cite{TlCoCl3inela}, namely the excitation is attributed to the large transverse magnetic fluctuation $S_{xx} ({\pmb Q}, \omega)$, while the spectral shape of the magnetic excitation observed in the SF scattering is different from that in the NSF scattering. We can expect that this difference is due to the longitudinal magnetic fluctuation $S_{zz} ({\pmb Q}, \omega)$. Note that this difference is not due to the phonon excitations because phonon excitations should be only observed in the NSF channel \cite{Moon}. \par
\begin{figure}[t]
\begin{center}
\includegraphics[width=85mm]{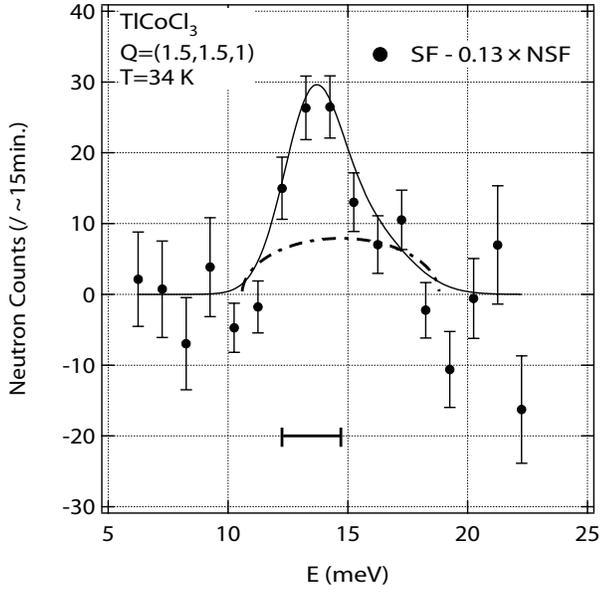}%
\end{center}
\caption{Energy dependence of the observed residual SF scattering intensity corresponding to the pure $S_{zz} ({\pmb Q}, \omega)$ contribution, as shown in eq. (\ref{H}), at ${\pmb Q}$ = (1.5, 1.5, 1). Here the energy dependent background is subtracted. In the estimation, we averaged the two neighboring data in order to reduce the statistical error. The dash-dotted line denotes the calculated residual SF scattering intensity based on the $S_{zz} ({\pmb Q}, \omega)$ spectrum of the IS theory without adjustable parameters (see text). The solid line is only a guide to the eyes. The horizontal bar indicates the calculated resolution width at $E=13.5$ meV. \label{Fig5}}
\end{figure}
In the following, we compare the obtained results at ${\pmb Q}=(1.5, 1.5, 1)$ with the IS theory. First, we consider the NSF scattering intensity. As mentioned above, $S_{xx} ({\pmb Q}, \omega)$ can be expressed as eq. (\ref{E}). Hence, at ${\pmb Q}=(1.5, 1.5, 1)$, the NSF scattering intensity can be expressed as follows:
\begin{equation}
\label{F}
I_{\rm NSF} = C \cdot S_{xx} ({\pmb Q}, \omega) \\
\end{equation}
with
\begin{equation}
\label{FF}
S_{xx} ({\pmb Q}, \omega) \simeq \left\{ \begin{array}{l}
                                        \frac{1}{32 \pi \epsilon^2 J^2} \sqrt{16 \epsilon^2 J^2 - \left( \omega - 2J \right)^2} \left[ 1 + 2 \epsilon \right. \\ 
                                        \hspace{0.5cm}  \left. - \frac{\omega - 2J}{J} \right] \hspace{0.5cm} \mbox{for} \hspace{0.5cm} |\omega - 2J| < 2|V| \\
                                        0 \hspace{0.5cm} \mbox{otherwise}
                                        \end{array} \right.                                      
\end{equation}
The NSF scattering intensity calculated using this $S_{xx} ({\pmb Q}, \omega)$ with the exchange constant $2J$ and the anistropy $\epsilon$ of 14.7 meV and 0.14, respectively, is plotted in Fig. \ref{Fig4} by the dash-dotted line. In order to fit the NSF scattering intensity obtained in the present measurements, the scale factor $C$ is estimated as 3000. As shown in Fig. \ref{Fig4}, the spectral shape is not very well reproduced. Note that the convolution of the resolution function is not considered in this comparison. However, this difference is not due to the convolution because the observed spectrum is sharper than the calculated one. The SF scattering intensity at ${\pmb Q}=(1.5, 1.5, 1)$ can be expressed as follows:
\begin{equation}
\label{G}
I_{\rm SF} = C \left( 0.13 S_{xx} ({\pmb Q}, \omega) + 0.87 S_{zz} ({\pmb Q}, \omega) \right)
\end{equation}
with
\begin{equation}
S_{zz} ({\pmb Q}, \omega) = \frac{\epsilon}{2 \pi J} \sqrt{1 - \left( \frac{\omega - 2J}{4 \epsilon J} \right)^2}
\end{equation}
and $S_{xx} ({\pmb Q}, \omega)$ shown in eq. (\ref{FF}). Hence the residual SF scattering intensity corresponding to the pure $S_{zz} ({\pmb Q}, \omega)$ contribution can be estimated from NSF and SF scattering intensities, namely
\begin{eqnarray}
\label{H}
I_{\rm res} &=& I_{\rm SF} - 0.13 I_{\rm NSF} \nonumber \\
    &=& C \cdot 0.87 S_{zz} ({\pmb Q}, \omega). 
\end{eqnarray}
Note that there are no adjustable parameters in this comparison because the scale factor $C$ is already estimated from the obtained NSF scattering intensity, as mentioned above. The observed and calculated residual SF scattering intensities after the subtraction of the linear energy dependent background are shown in Fig. \ref{Fig5}. For the estimation of $I_{\rm res}$, we averaged the two neighboring data in order to reduce the statistical error. This procedure can be considered to be valid because the experimental resolution in the energy scan is approximately 2.5 meV, as shown in Fig. \ref{Fig5}. The intensity of the observed longitudinal fluctuation $S_{zz} ({\pmb Q}, \omega)$ in the vicinity of $E=2J=14.7$ meV is quite well reproduced by the IS theory. Note that the intensity of the longitudinal fluctuation $S_{zz} ({\pmb Q}, \omega)$ is much smaller than the transverse fluctuation $S_{xx} ({\pmb Q}, \omega)$ in accord with the theoretical prediction. The spectral shapes of the experimentally observed $S_{zz} ({\pmb Q}, \omega)$ and $S_{xx} ({\pmb Q}, \omega)$ do not quite well correspond to those calculated by the IS theory. There are some discussions, in which this difference is attributed to the additional terms, such as the exchange mixing \cite{Goff,Nagler} and the next-nearest-neighbor ferromagnetic interaction \cite{Matsubara,Shiba2}, which are not considered in the IS theory. The comparison between the calculated spectra including the additional terms and the observed ones in the present experiments is needed.

\section{Conclusion}

We have presented the results of the polarized neutron inelastic scattering measurements in the quasi-1D Ising-like antiferromagnet TlCoCl$_3$. We clearly identified the longitudinal magnetic fluctuation $S_{zz} ({\pmb Q}, \omega)$ for the spin-wave excitation continuum at $\omega \sim 2J$, which has not been observed in the unpolarized neutron inelastic scattering experiments of the quasi-1D Ising-like antiferromagnets CsCoCl$_3$ and TlCoCl$_3$ so far, together with the transverse magnetic fluctuation $S_{xx} ({\pmb Q}, \omega)$. It was found that the observed longitudinal and transverse magnetic fluctuation spectra at characteristic wave vectors semi-quantitatively correspond to those calculated by the perturbation theory from the pure Ising limit, as shown in Figs. \ref{Fig2} through \ref{Fig5}.

\section*{Acknowledgement}
 
We acknowledge M. Takeda and Y. Shimojo for technical supports. This work was supported by the Saneyoshi Scholarship Foundation and the Kurata Memorial Hitachi Science and Technology Foundation. \par


\begin{thebibliography}{99}
\bibitem{Yoshizawa} H. Yoshizawa, K. Hirakawa, S. K. Satija and G. Shirane: Phys. Rev. B {\bf 23} (1981) 2298.
\bibitem{Nagler} S. E. Nagler, W. J. L. Buyers, R. L. Armstrong and B. Briat: Phys. Rev. B {\bf 27} (1983) 1784.
\bibitem{Villain} J. Villain: Physica B {\bf 79} (1975) 1.
\bibitem{Nishiwaki1} Y. Nishiwaki, K. Iio and T. Mitsui: J. Phys. Soc. Jpn. {\bf 72} (2003) 2608.
\bibitem{Nishiwaki2} Y. Nishiwaki, T. Kato, Y. Oohara, A. Oosawa, N. Todoroki, N. Igawa, Y. Ishii and K. Iio: J. Phys. Soc. Jpn {\bf 75} (2006) 034707.
\bibitem{TlCoCl3inela} A. Oosawa, Y. Nishiwaki, T. Kato and K. Kakurai: J. Phys. Soc. Jpn. {\bf 75} (2006) 015002.
\bibitem{Ishimura} N. Ishimura and H. Shiba: Progr. Theor. Phys. {\bf 63} (1980) 743.
\bibitem{Moon} R. M. Moon, T. Riste and W. C. Koehler: Phys. Rev. {\bf 181} (1969) 920.
\bibitem{Goff} J. P. Goff, D. A. Tennant and S. E. Nagler: Phys. Rev. B {\bf 52} (1995) 15992.
\bibitem{Matsubara} F. Matsubara and S. Inawashiro: Phys. Rev. B {\bf 43} (1991) 796.
\bibitem{Shiba2} H. Shiba, Y. Ueda, K. Okunishi, S. Kimura and K. Kindo: J. Phys. Soc. Jpn. {\bf 72} (2003) 2326.
\end{thebibliography}
\end{document}